\documentclass{aastex}   
\usepackage{emulateapj5}
\usepackage{natbib}  
\shorttitle{Eddington limit and radiative transfer in inhomogeneous atmospheres}
\shortauthors{M. Ruszkowski and M.C. Begelman}

\received{2002 April 15}
\begin{document}

\title{Eddington limit and radiative transfer in highly  inhomogeneous atmospheres}

\author{Mateusz Ruszkowski and Mitchell C. Begelman\altaffilmark{1}}
\affil{JILA, Campus Box 440, University of Colorado, Boulder CO 80309-0440}
\email{mr@quixote.colorado.edu; mitch@jila.colorado.edu}

\altaffiltext{1}{Also at Department of Astrophysical and Planetary
Sciences, University of Colorado}

\begin{abstract}
Radiation dominated accretion disks are likely to be subject to 
the ``photon bubble'' instability, which may lead to strong
density inhomogeneities on scales much shorter than the disk scale height. 
Such disks -- and magnetized, radiation-dominated atmospheres in
general -- 
could radiate well above the Eddington limit without being disrupted. 
When density contrasts become large over distances of order the photon mean
free path, radiative transfer cannot be described adequately 
using either the standard diffusion approximation or existing
prescriptions for flux-limited diffusion. Using analytical and Monte Carlo 
techniques, we consider the effects of strong density gradients 
deep within radiation- and scattering-dominated atmospheres.
We find that radiation viscosity -- i.e., the off-diagonal elements of the
radiation stress tensor -- has an important effect on radiative
transfer under such conditions.
We compare analytical and numerical results in the specific case of a
plane-parallel density wave structure and calculate Eddington enhancement 
factors due to the porosity of the atmosphere.
Our results can be applied to the study of dynamical coupling between 
radiation forces and density
inhomogeneities in radiation dominated accretion disks in two or three
dimensions.
\end{abstract}


\section{Introduction}
Radiation-dominated atmospheres of accretion disks and massive stars 
permeated
by a moderately strong magnetic field are susceptible to the ``photon
bubble'' instability \citep{ar92,ga98,be01}. In this process, 
high density regions tend to be pulled downward along the field lines 
and low density regions are pushed upward by radiation forces. The gas
elements accelerated by radiation forces enter density maxima, where
radiation forces decrease, and then progress downward again completing the 
cycle. Because subsequent acceleration episodes are increasingly large and 
magnetic tension prevents high density regions from spreading sideways, 
the density contrast increases.
This leads to 
density inhomogeneities on scales shorter than the characteristic scale 
height of the accretion disk or stellar atmosphere. 
Under such conditions, radiation tends to bypass high density
regions and travel more freely through tenuous ones. 
If the low- and high-density regions are dynamically coupled and most of
the mass is in the high density phase, then the flux
necessary to support the atmosphere against gravity can exceed
Eddington limit. Photon bubble instability may be
applicable to objects suspected of having
super-Eddington luminosities, such as ``ultraluminous X-ray sources''
\citep{be02} and narrow-line Seyfert 1 galaxies \citep{pu02}.
It has also been argued that the instability may occur in simple
non-magnetized Thomson atmospheres as they approach the Eddington limit \citep{sh01b}.
The origin of this instability and the statistical properties of the
inhomogeneities, 
such as their size relative to the size of the atmosphere, are different
than that predicted by the magnetic photon bubble instability.
This mechanism has been invoked
to explain the discrepancy between relatively high outburst luminosities
and relatively low outflow velocities in novae \citep{sh98,sh00,sh01a}.\\
\indent
Radiative transfer in a medium where parts of the gas are optically thin,
such as the upper regions of atmospheres in particular, cannot be treated
using the standard diffusion approximation. The application of this
approximation in the optically thin regime may lead to the radiation
propagation rate exceeding the free-streaming rate $|\mathbf{F}|=cu$, where
$\mathbf{F}$ and $u$ are the radiation flux and energy density,
respectively. Previous work on radiative transfer under such conditions 
focused on the development of flux-limited diffusion approximations
(e.g., \citet{le81,me91,an92}). A serious limitation of these methods is 
that they are applicable only in cases 
where the angular distribution of the specific intensity is a
slowly varying function of space and time.\\
\indent
In this paper, we focus on radiative transfer deep within atmospheres
(i.e., at high optical depth), where
the effects of flux-limited diffusion are less important. 
But we consider the case where large density gradients 
-- on scales of order the photon mean free path --
lead to rapid fluctuations in the angular distribution function.
It is easy to see why the standard diffusion approximations (with or without
flux-limiting) fail under these circumstances.
The standard diffusion approach predicts that the flux responds
instantaneously to local changes in density. But in reality the flux can
only respond to inhomogeneities provided that 
the density changes on scales much larger
than the photon mean free path. When density inhomogeneities are optically
thin, radiation does not ``see'' any density contrasts.
Using a simple ansatz, we have derived a modified diffusion equation and
 show that the effects of ``photon viscosity''
(i.e., off-diagonal elements of the radiation stress tensor) 
play an important role in
radiative transfer in this case. Using analytical and Monte Carlo
techniques we find that our analytical approach is much more accurate than 
 standard diffusion or multi-stream approximations. 
Our results can be applied to the study of dynamical coupling between
 radiation forces and density
inhomogeneities in radiation-dominated accretion disks.\\
\indent
The structure of this paper is as follows. In section 2 we discuss 
three-dimensional radiative transfer and show that the ``photon viscosity''
terms are important. 
In section 3 we constrain ourselves to the case of periodic, planar 
density waves embedded deep within an atmosphere 
and derive the corresponding Eddington enhancement factors under the 
assumption of global dynamical equilibrium.
In section 4 we compare analytical formulas with Monte Carlo calculations and
discuss our results. We propose that our method could be incorporated into
existing radiation hydrodynamics codes such as the RHD module for ZEUS
developed by \citet{tu01}.

\section{Radiative transfer in an inhomogeneous atmosphere}
We consider radiative transfer within infinite highly inhomogeneous
atmospheres where the effects of flux-limited diffusion are very much
reduced. Moreover, in a porous atmosphere, the 
radiation diffusion timescale is
most likely much shorter than the characteristic gas dynamical timescale, 
i.e., radiation is probably not trapped by the motion of gaseous 
 inhomogeneities
\citep{be01}. 
Therefore, for the purpose of calculating radiative transfer, we can 
neglect the time-dependence and motion of the gas distribution.
We approximate the intensity distribution as

\begin{equation}
I(\mathbf{x},\hat{\Omega})=I_{0}(\mathbf{x},\hat{\Omega})+
\frac{3}{4\pi}\hat{\Omega}\cdot\mathbf{F}(\mathbf{x}),
\end{equation}

\noindent
where $\mathbf{F}$ is the local flux vector, $\hat{\Omega}$ is the
directional unit vector, and $\mathbf{x}$ is the position vector. 
This approximation means that the first nontrivial correction to 
intensity $I_{0}(\mathbf{x},\hat\Omega)$ is symmetric with respect to the
direction defined by the {\it local} flux. This has to be contrasted with
a familiar case of a plane-parallel atmosphere. In such a case, it is
customary to assume that the directional distribution of intensity 
$I(\mathbf{x},\hat\Omega)$ is symmetric relative to the vertical direction
which coincides with the direction of the flux vector. However, in the case
of highly inhomogeneous atmosphere, radiation bypasses denser regions and
the flux vector can rapidly change its orientation and only the
volume-averaged flux is vertical. Thus, in our approach, the local
``symmetry axis'' is changing direction throughout the atmosphere and no
global symmetry is required.
We also
assume that all the odd moments of $I_{0}$ vanish, i.e., 
$\int\hat{\Omega}_{i}I_{0}d\Omega=\int\hat{\Omega}_{i}\hat{\Omega}_{j}
\hat{\Omega}_{k}I_{0}d\Omega=0$, but otherwise make no assumptions about the
directional dependence of $I_{0}$. 
The equation of radiative transfer for a 3D scattering atmosphere reads

\begin{equation}
\frac{1}{\sigma}\hat{\Omega}\cdot\nabla I=-I+\frac{1}{4\pi}\int Id\Omega,
\end{equation}

\noindent
where $\sigma=\rho\kappa$ is the scattering coefficient. Taking zeroth and
first moments of eq. (2) we obtain

\begin{equation}
\nabla\cdot\mathbf{F}=0
\end{equation}
and
\begin{equation}
\mathbf{F}=-\frac{c}{\sigma}\nabla\cdot\mathbf{T} ,
\end{equation}

\noindent
where $\mathbf{T}$ is the radiation stress tensor, the components 
of which are 
\begin{equation}
T_{ij}=\frac{1}{c}\int\hat{\Omega}_{i}\hat{\Omega}_{j}I_{0}d\Omega .
\end{equation}

\noindent
The closure relation for $T_{ij}$
in terms of $F_{i}$ and $J\equiv \frac{1}{4\pi}\int Id\Omega$ 
can be obtained by calculating the second moment 
of eq. (2), assuming the form of the intensity given by eq. (1). This leads to
the equation for the radiation stress tensor:

\begin{equation}
T_{ij} =\left\{ \begin{array}{ll}
\frac{u}{3}-\frac{1}{5\sigma c}\left(\frac{\partial F_{i}}{\partial x^{j}}+\frac{\partial F_{j}}{\partial x^{i}}\right)
& \mbox{for}\; i=j\\
-\frac{1}{10\sigma c}\left(\frac{\partial F_{i}}{\partial x^{j}}+\frac{\partial F_{j}}{\partial x^{i}}\right) & \mbox{for}\; i\neq j
\end{array}\right.
\end{equation}

\noindent
where $u=4\pi J/c$ is the energy density and $J$ is the mean intensity.
Note that (i) the diagonal terms $T_{ii}$ may be different from one
another, and (ii) 
the off-diagonal elements of the radiation stress tensor do not 
vanish. The first point implies that this approach incorporates variable 
Eddington factors. The off-diagonal 
elements are responsible for ``photon viscosity'' and are non-zero even
though bulk gas motions in the atmosphere were assumed to be negligible.
This is because the ``photon fluid'' is moving through the gas, and may
exert shear stresses.
Substituting eq. (6) into eq. (4) and using eq. (3), we obtain:

\begin{eqnarray}
F_{i} & = & -\frac{c}{3\sigma}\frac{\partial u}{\partial x^{i}}+
\frac{1}{10\sigma^{2}}\left[\frac{\partial^{2}F_{i}}{\partial x_{j}x^{j}}
+2\frac{\partial^{2}F_{i}}{\partial x_{i}^{2}}-\right.\nonumber \\
& & \left.\frac{1}{\sigma}\frac{\partial\sigma}{\partial x_{j}}
\left(\frac{\partial F_{i}}{\partial x^{j}}+\frac{\partial F_{j}}{\partial
x^{i}}\right)-\frac{2}{\sigma}\frac{\partial\sigma}{\partial x^{i}}
\frac{\partial F_{i}}{\partial x_{i}}\right] .
\end{eqnarray}

\noindent
In the above equation, summation is imposed {\it only} over repeated $j$ 
indices. Equation (7) and equation (3) are the governing equations of
radiative transfer in our approximation.

\section{2D density inhomogeneities}
In order to illustrate our method, we now consider the simplified case 
of a plane-parallel wave density pattern (see Fig. 1). 
The scattering
coefficient is assumed to be only a function of the distance $\xi$ 
perpendicular to the slabs, i.e., 
$\sigma =\rho\kappa =\sigma(\xi)$ with $\xi =\mu
x+(1-\mu^{2})^{1/2}z$, where $\mu =\cos\psi$. We also assume 
that $\partial/\partial y=0$ and that the components of flux depend only 
on $\xi$. We consider an atmosphere which is in {\it global} 
(i.e., volume-averaged)
hydrostatic equilibrium, where radiation pressure balances gravity
$-g\hat{z}$. This implies that the gradient of energy density must be of
the form

\begin{equation}
\nabla u=u'(\xi)\nabla\xi -\frac{3\langle\sigma\rangle}{c}F_{\rm Edd}\hat{z},
\end{equation}

\noindent
where a prime denotes differentiation with respect to $\xi$, 
$\langle\sigma\rangle$ is the volume average of the scattering coefficient,
and $F_{\rm Edd}=gc/\kappa$ is the Eddington flux. From equation (7) and 
equation (3) we obtain:

\begin{equation}
\mu
F_{z}-(1-\mu^{2})^{1/2}F_{x}=\mu\frac{\langle\sigma\rangle}{\sigma}F_{\rm Edd}+
\frac{1}{\alpha^{2}\sigma\mu}\left(\frac{F_{z}'}{\sigma}\right)' ,
\end{equation}

\noindent
where $\alpha^{2}(\mu)=10/(1+4\mu^{2}-4\mu^{4})$. Integrating eq. (3)
over $\xi$ we obtain

\begin{equation}
\mu F_{x}+(1-\mu^{2})^{1/2}F_{z}=const\equiv (1-\mu^{2})^{1/2}F_{0} ,
\end{equation}

\noindent
where $F_{0}$ is the integration constant.
Using eq. (10) to eliminate $F_{x}$ from eq. (9) we get

\begin{equation}
F_{z}=(1-\mu^{2})F_{0}+\mu^{2}\frac{\langle\sigma\rangle}{\sigma}F_{\rm Edd}+
\frac{1}{\alpha^{2}\sigma}\left(\frac{F_{z}'}{\sigma}\right)'.
\end{equation}

\noindent
Multiplying equation (11) by $\sigma$ and then volume-averaging it and
demanding that the solution be bounded, we have

\begin{equation}
\langle\sigma\rangle F_{\rm Edd}=\langle\sigma F_{z}\rangle =
(1-\mu^{2})\langle\sigma\rangle F_{0}+\mu^{2}\langle\sigma\rangle F_{\rm
Edd} ,
\end{equation}

\noindent
where the first equality in equation (12) comes from the requirement that 
the atmosphere be in global hydrostatic equilibrium. Therefore,
$F_{0}=F_{\rm Edd}$ and the final equation for the vertical flux is

\begin{equation}
F_{z}=\left(1-\mu^{2}+\mu^{2}\frac{\langle\sigma\rangle}{\sigma}\right)F_{\rm
Edd}+
\frac{1}{\alpha^{2}\sigma}\left(\frac{F_{z}'}{\sigma}\right)'.
\end{equation}

\subsection{Eddington enhancement factor}
Using equation (13) we can now calculate the Eddington enhancement 
factor $l\equiv \langle F_{z}/F_{\rm Edd}\rangle$. 
We can simplify and non-dimensionalize equation (13) by defining

\begin{equation}
f\equiv\frac{F_{z}}{F_{\rm Edd}}-(1-\mu^{2}) .
\end{equation} 

\noindent
Thus, the Eddington factor is

\begin{equation}
l\equiv (1-\mu^{2})+\langle f\rangle ,
\end{equation} 

\noindent
where $\langle f\rangle$ is the volume average of $f$.
Defining optical depth $d\tau =\sigma d\xi$, letting the prime now
denote differentiation with respect to $\tau$, and normalizing the scattering
coefficient to its mean value
(i.e. $\sigma\rightarrow\sigma/\langle\sigma\rangle$), we obtain a
simplified form of equation (13)

\begin{equation}
f''-\alpha^{2}f=-\frac{\alpha^{2}\mu^{2}}{\sigma} .
\end{equation}

\noindent
We now consider a periodic slab model in which the scattering coefficient is
given by:

\begin{equation}
\sigma =\left\{ \begin{array}{ll}
             \sigma_{1} & \mbox{if $-\tau_{1}<\tau<0$, region 1}\\
             \sigma_{2} & \mbox{if $\; \; \; 0<\tau<\tau_{2}$, region 2 .}
\end{array}\right.
\end{equation}

\noindent
The solutions to equation (16) in regions 1 and 2 are

\begin{equation}
f_{1,2}(\tau)=a_{1,2}\sinh(\alpha\tau)+b_{1,2}\cosh(\alpha\tau)+
\frac{\mu^{2}}{\sigma_{1,2}},
\end{equation}

\noindent
where $a_{1,2}$ and $b_{1,2}$ are the integration constants. Functions $f$
and $f'$ have to be continuous across each slab boundary. Therefore the 
matching conditions are:

\begin{eqnarray}
\;\;\; f_{1}(0)=f_{2}(0), & \;\;\; f_{1}'(0)=f_{2}'(0)   \\
f_{1}(-\tau_{1})=f_{2}(\tau_{2}), & f_{1}'(-\tau_{1})=f_{2}'(\tau_{2}) .
\end{eqnarray} 

\noindent
Using the above matching conditions to derive the integration constants 
$a_{1,2}$ and $b_{1,2}$ and then volume averaging the solution for $f$, we
obtain the expression for $\langle f\rangle$

\begin{equation}
\langle f\rangle =\frac{\mu^{2}\langle\sigma\rangle}{\xi_{1}+\xi_{2}}
\left[\frac{\xi_{1}}{\sigma_{1}}+\frac{\xi_{2}}{\sigma_{2}}-
\frac{2}{\alpha}\frac{(\Delta\sigma)^{2}}{\sigma_{1}^{2}\sigma_{2}^{2}}
\frac{\sinh x\sinh y}{\sinh (x+y)}\right] ,
\end{equation}

\noindent
where $x=\alpha\tau_{1}/2$, $y=\alpha\tau_{2}/2$, $\Delta\sigma
=\sigma_{2}-\sigma_{1}$, and $\xi_{1,2}=\tau_{1,2}/\sigma_{1,2}$.
The Eddington enhancement factor $l$ follows directly from equations (21) 
and (15).

\section{Comparison with Monte Carlo simulations}
We now compare our analytical results with Monte Carlo
simulations. The setup of the numerical experiment was as follows. We
instantaneously injected a large number of photons in the equatorial plane
(i.e., $z=0$) of a very flat, three-dimensional box (i.e., $z\in
[-z_{o},z_{o}]$; $x,y\in [-w_{o},w_{o}]$, where $z_{o}\ll
w_{o}$). Densities and height of the computational box were chosen 
in such a
way as to assure that the optical thickness in the vertical 
direction would 
always be large throughout the box. The initial photon angular
distribution was uniform. Although we included the anisotropy due
to the Thomson scattering cross section, we found this effect to have no
 influence on our final results. 
We followed the trajectories of all photons and calculated photon 
travel times
between scatterings. Momentum transfer for every scattering was calculated 
using the method of weights \citep{po83}. We then computed the force 
exerted on the atmosphere as a function of the time delay
following the instantaneous photon injection. 
Of course, as photons diffuse out of the atmosphere, the force exerted 
on the gas gradually declines. Therefore, the total force,
corresponding to a continuous photon flux, was
calculated by superposing many such time-dependent force distributions 
due to groups of 
photons injected (instantaneously) at uniform time intervals.
The total force exerted on the atmosphere was characterized by a gradual
increase with time followed by a flat maximum.
The Eddington enhancement factor is then given by the
ratio of the 
``saturated'' total force (i.e., total, constant force at late times)
exerted on a homogeneous atmosphere, 
to the total force acting on an inhomogeneous atmosphere characterized by
the same mean density. 
Note that the Eddington enhancement factor, defined in this way, can also be
interpreted as {\it the ratio of 
fluxes necessary to 
exert  the same amount of force on an inhomogeneous atmosphere
as on the corresponding homogeneous atmosphere}. This ratio is the same
even if the homogeneous atmosphere is sub-Eddington, i.e., is only
partially supported against gravity by radiation.

Figure 2 shows the Eddington enhancement factor for variable inclination of 
slabs (upper panels) and for changing density contrast of vertical slabs 
(lower panels). In all cases, the Thomson depth of the high density slabs 
$\tau_{h}$ is constant but
the optical depth across
 low density regions increases from values $\tau_{l}<1$ to
$\tau_{l}>1$ from left to right (see caption of Fig. 2 for more
details). As expected, the Eddington factor increases as the slabs rotate
toward the vertical direction because the atmosphere effectively becomes
more porous (see upper panels). 
When the slabs are vertical, the flux enhancement factor increases 
as the density contrast $\sigma_{h}/\sigma_{l}$ becomes larger for constant
mean density. This is due to the fact that the volume filling factor of
the high-density gas decreases while that of the low-density gas 
increases, but the respective masses of the two density phases remain the
same. Therefore, the mean, volume-weighted, flux is

\begin{equation}
\langle F\rangle =(1-f_{v})F_{l}+f_{v}F_{h}\approx F_{l},
\end{equation}

\noindent
where $f_{v}$ is the volume filling factor of the dense gas and 
$F_{l}$ and $F_{h}$ are the fluxes propagating through tenuous and
dense regions, respectively.
As the density contrast increases and $f_{v}$ decreases, radiation tends
to ``flow'' primarily through the low density channels and, therefore, more
flux is necessary to exert the same total force as in the homogeneous case
because radiation interacts less efficiently with tenuous gas.
Quantitatively, in the diffusion limit, we have \citep{sh98}

\begin{equation}
l=\langle\rho\rangle\left\langle\frac{1}{\rho}\right\rangle=
\left(\frac{\xi_{l}}{\sigma_{l}}+\frac{\xi_{h}}{\sigma_{h}}\right)
\frac{\langle\sigma\rangle}{\xi_{l}+\xi_{h}} .
\end{equation}

\noindent
When most volume is in the low-density phase but most mass is in the 
high-density phase
(i.e., $f_{v}\rightarrow 0$ or $\xi_{h}/\xi_{l}\rightarrow 0$), then
$l\approx f_{v}\rho_{h}/\rho_{l}$ if $f_{v}\gg\rho_{l}/\rho_{h}$. This
qualitatively explains why $l$ decreases with $\tau_{l}\ge 1$ at constant
density contrast $\sigma_{h}/\sigma_{l}$ (cf. third and fourth columns on
Fig. 2). At small optical depth $\tau_{l}$, equation (23) would lead to
very inaccurate answers. For example, equation (23) predicts
$l\sim 23$ for vertical slabs with $\tau_{l}=0.1$ and
$\sigma_{h}/\sigma_{l}=100$, compared to the actual value $l\sim 5$ and 
our analytic result $\sim 4$ from eq. (21) (cf. lower left panel). 
This discrepancy is 
due largely to neglect of the anisotropy of the radiation field, whereas 
our approach gives much more accurate results even in such an extreme case.
Moreover, note that the ``anisotropy term'' in our expression for the flux 
enhancement factor, which is proportional to $(\Delta\sigma)^{2}
=(\sigma_{h}-\sigma_{l})^{2}$, vanishes for large Thomson depths and thus
equations (21) and (15) reduce
 to equation (23) in the diffusion limit. We also
considered ``multi-stream'' approximation schemes in order to account 
for the
radiation anisotropy, but found the ``intensity moment'' 
approach developed here to be in 
significantly better agreement with Monte Carlo simulations.

\section{Summary}
We have considered radiative transfer deep within extremely 
inhomogeneous atmospheres, and have
demonstrated that, under such conditions, radiation viscosity 
-- i.e., the off-diagonal elements of the radiation stress tensor -- 
plays an
important role. Our approach is significantly more accurate than 
approaches based on the diffusion equation and multi-stream approximation. 
The technique developed here can be applied to the
nonlinear evolution of radiation-driven instabilities in 
accretion disks.
In particular, it can be used to study the dynamical coupling 
of matter and
radiation in order to determine the characteristic length scales 
and density
contrasts arising from ``photon bubble'' instability. This, in turn, 
will permit
a self-consistent determination of the magnitude of the Eddington 
enhancement
factor in radiation-dominated accretion disks. We also suggest that 
our method
could be incorporated into radiation hydrodynamics codes
such as the RHD module for ZEUS \citep{tu01}. \\

This work was supported in part by NSF grant AST--9876887.
We thank Nir Shaviv and Neal Turner for comments on the manuscript.

\clearpage

\begin{figure}
\plotone{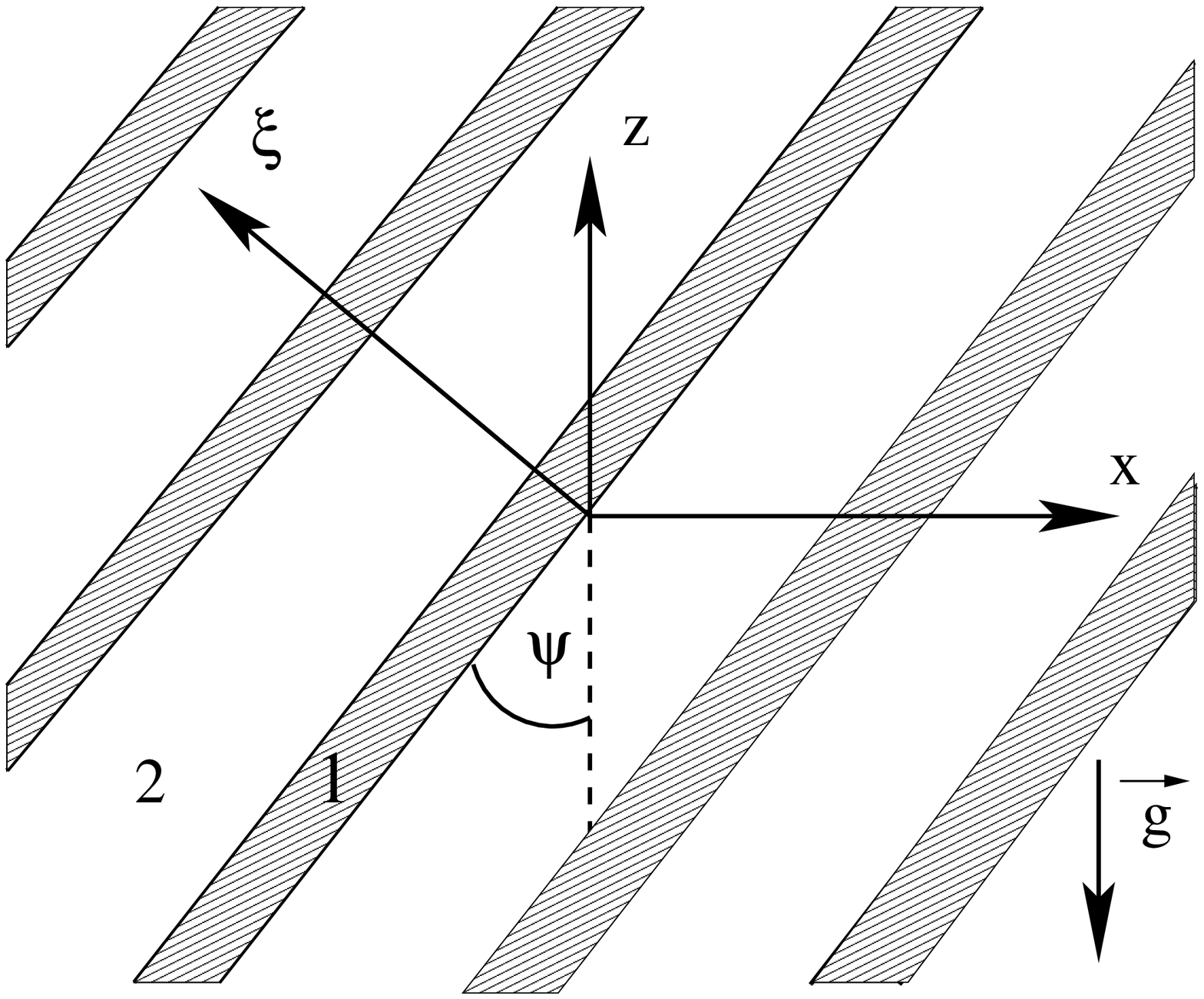}
\caption{Density structure of an inhomogeneous atmosphere. Shaded and
unshaded zones denote higher and lower density regions, respectively. 
The region shown is much smaller than the overall radiation pressure scale
height $\left|\frac{u}{\nabla u}\right|$.}
\end{figure}

\clearpage

\begin{figure}
\plotone{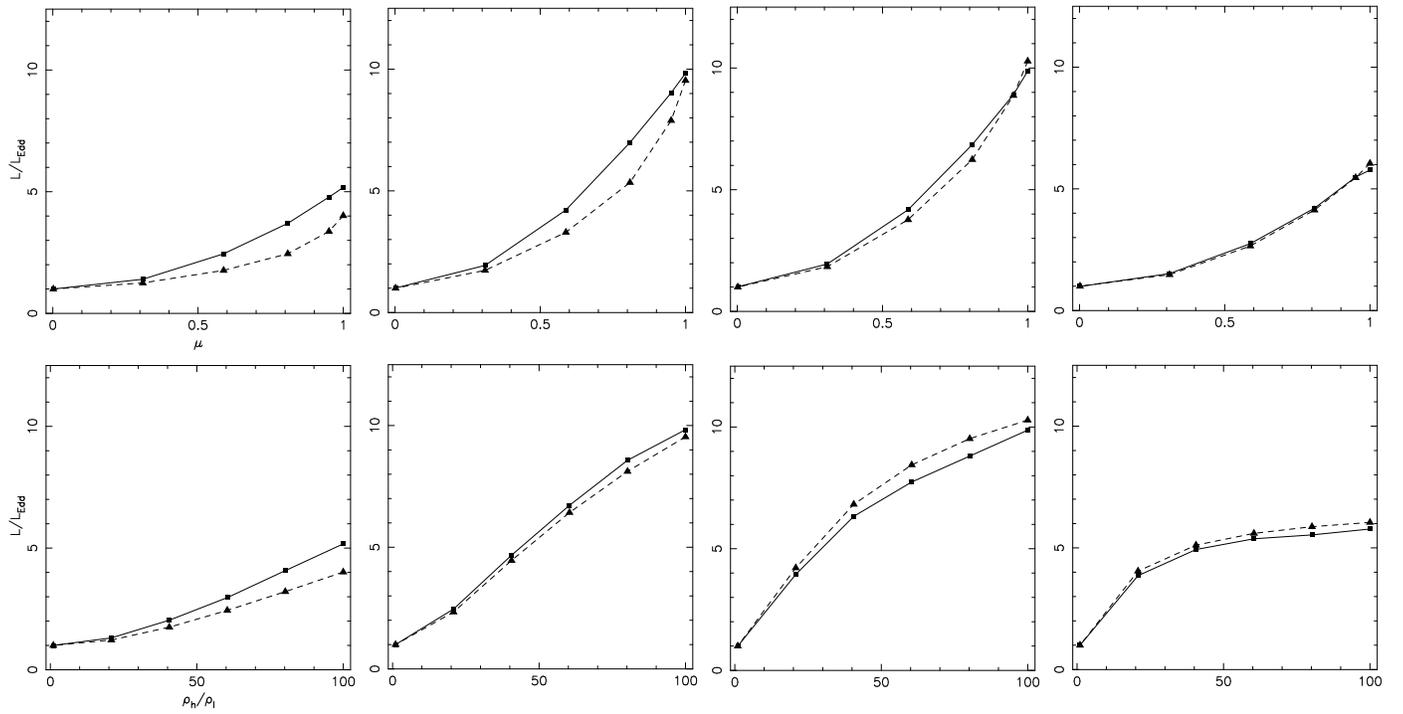}
\caption{Eddington enhancement factors $L/L_{\rm Edd}$ for $\tau_{h}=20$ and
$\tau_{l}=0.10,\; 0.33,\; 1.00,\; 3.00$ (from left to right column, 
respectively). Upper panels show $L/L_{\rm Edd}$ as a function of $\mu
=\cos\psi$ for scattering coefficients of low and high density regions
equal to 
$\sigma_{l}=10$ and $\sigma_{h}=1000$, respectively. Lower panels show 
 $L/L_{\rm Edd}$ as a function of $\sigma_{h}/\sigma_{l}$ for vertical
slabs. Dashed lines show analytical predictions from equations (15) and
(21), while solid lines show results from Monte Carlo simulations.}
\end{figure}

\end{document}